\documentclass{article}
\usepackage[utf8]{inputenc}
\usepackage{amsmath}
\usepackage{color}
\usepackage{amsmath, amsfonts, amsthm, amssymb}
\usepackage{mathtools}
\usepackage{commath}
\usepackage{verbatim}
\usepackage[toc,page]{appendix}
\usepackage{stmaryrd}
\usepackage{fancyhdr}
\usepackage{bbm}
\usepackage{graphicx}
\usepackage{graphics}
\usepackage{tikz}
\usepackage{caption}
\usepackage{subcaption}
\usepackage{booktabs}
\usepackage{array}
\usepackage{comment}
\usepackage{todonotes}
\usepackage{tcolorbox}
\usepackage{float}
\usepackage{algorithm}
\usepackage[noend]{algpseudocode}
\usepackage{tikz}
\usepackage{multirow,array}
\usepackage{color}

\usepackage[backend=bibtex, natbib=true, bibstyle=numeric, citestyle=numeric, doi=true, sorting=nyt, giveninits]{biblatex}
\usepackage{listings}
\usepackage{xcolor}
\newfam\msbfam

\usepackage[affil-it]{authblk}

\theoremstyle{definition}

\title{On the derivation of the renewal equation from an age-dependent branching process: an epidemic modelling perspective}

\author{Swapnil Mishra$^{*,1,\dagger}$, Tresnia Berah$^{*,1}$, Thomas A. Mellan$^{*,1}$,  H. Juliette T. Unwin$^{1}$, Michaela A Vollmer$^{1}$, Kris V Parag$^{1}$, Axel Gandy$^{2}$, Seth Flaxman$^{2}$, and Samir Bhatt$^\dagger$}
\affil{$^1$ Department of Infectious Disease Epidemiology, Imperial College London}
\affil{$^2$ Department of Mathematics, Imperial College London}
\affil{$^*$ Joint Authorship}
\affil{$^\dagger$ Corresponding Author s.bhatt@imperial.ac.uk, \\s.mishra@imperial.ac.uk}
\date{}
\begin{document}

\maketitle

\begin{abstract}
 Renewal processes are a popular approach used in modelling infectious disease outbreaks. In a renewal process, previous infections give rise to future infections. However, while this formulation seems sensible, its application to infectious disease can be difficult to justify from first principles. It has been shown from the seminal work of Bellman and Harris \cite{Bellman1948} that the renewal equation arises as the expectation of an age-dependent branching process. In this paper we provide a detailed derivation of the original Bellman Harris process. We introduce generalisations, that allow for time-varying reproduction numbers and the accounting of exogenous events, such as importations. We show how inference on the renewal equation is easy to accomplish within a Bayesian hierarchical framework. Using off the shelf MCMC packages, we fit to South Korea COVID-19 case data to estimate reproduction numbers and importations. Our derivation provides the mathematical fundamentals and assumptions underpinning the use of the renewal equation for modelling outbreaks. 
\end{abstract}

\section{Introduction}
Mathematical descriptions of infectious disease outbreaks are fundamental to forecasting and simulating the dynamics of epidemics, as well as to understanding the mechanics of how transmission occurs. One popular approach to modelling infectious disease outbreaks is founded on renewal processes and the renewal equation. A renewal process generalises a Poisson process to allow for arbitrary (instead of exponential) holding or event waiting times. From an infectious disease perspective, these holding times model how new infections are generated across an epidemic. The expected number of newly infections at some time is then given by the renewal equation \cite{Feller1941}. While this formulation seems sensible, its application to infectious disease can be difficult to justify from first principles since it does not propose a formal mechanism for the generation of infectious events. 

In the seminal work of \cite{Kermack1927}, Kermack and McKendrick studied the number and distribution of cases of an infectious disease as it is progressed through a population over time. They constructed classes, called compartments, and modelled the propagation of infectious disease via interactions among these compartments. The result is the popular susceptible-infected-recovered (SIR) model, variants of which are widely used in epidemiology. SIR models provide an intuitive mechanism for understanding disease transmission, and in the original derivation of \cite{Kermack1927}, they were found to be  similar to the Volterra equation \cite{Polyanin1998}. The Volterra equation is equal to the renewal equation when a convolution operator is applied. This link has been used to justify the role of renewal equations in modelling epidemic processes \cite{Fraser2007,Cori2013,Nouvellet2018,Cauchemez2016}. However, the connection between the renewal equation and compartmental models is neither intuitive nor simple and the consequences of the differing assumptions behind both mathematical descriptions can be difficult to assess, especially when the number of compartments or renewal parameters becomes large \cite{Champredon2018,Parag703751}.

The renewal equation may also be derived by treating every new infection event as a birth in the infected population. The Euler-Lotka equation from ecology \cite{Lotka1907}, which tracks the numbers of females in an age-structured population, can then be shown to yield the standard renewal equation of epidemiology \cite{Fraser2007}. While this derivation is simple and intuitive, it is somewhat limited. Specifically, it remains unclear how to properly incorporate repeated, imported infections, which can significantly impact the time-course of an epidemic. A stronger and more transparent link between the physical process underlying an infectious outbreak and the formulation of the renewal model is therefore warranted.

Bellman and Harris elegantly captured this underlying infection mechanism by formulating an age-dependent branching process \cite{Bellman1948}. Branching processes describe how individuals stochastically propagate their numbers over time. In epidemiology, age-insensitive branching processes, such as the fundamental Galton-Watson process, which discretise the propagation process into generations, have provided tractable yet intuitive ways of modelling the spread of an infectious disease \cite{Bartoszynski1967,Getz2006}. However, these approaches, while useful, lack the flexibility and realism of the more general Bellman-Harris processes \cite{Bellman1948,Bellman1952}. Age-dependence allows for the variable time between exposure to a pathogen and the onset of symptoms to be properly modelled and provides a framework for encode useful information on the biology of the infecting pathogen, such as incubation periods and non-monotonic infectiousness. Given these advantages, it is surprising that only a few epidemiological studies have considered Bellman-Harris approaches \cite{bharucha-reid1956}.

To understand the Bellman-Harris process we consider a homogeneous or well-mixed population, in which members can randomly infect one another. Let $t\in \mathbb{R}^+$, be a positive real number representing time. We introduce a positive random variable $\tau\in \mathbb{R}^+$ with probability distribution $g(\tau)$ and cumulative distribution $G(\tau) = \int_{t=0}^\tau g(t) dt$. After some random period $\tau,$ an infected individual can infect $n\in\mathbb{I}^+$ (positive integers) other individuals with probability $q_n$. We are interested in characterising the number of newly infected individuals at time $t$, which we denote $Z(t)$. However, as each time trajectory of $Z(t)$ is one possible reality (or sample-path) from the epidemic process, we more broadly aim to calculate and subsequently model, the average number of new infection events at time $t$, $E[Z(t)]$. 

Direct calculation of $E[Z(t)]$ requires the manipulation of a generally intractable integral. Consequently, we adopt a generating function based approach, which allows the expectation to be obtained through derivatives. Generating functions are extremely useful but often hard to understand. They are perhaps best described by Pólya: \emph{A generating function is a device somewhat similar to a bag. Instead of carrying many little objects detachedly, which could be embarrassing, we put them all in a bag, and then we have only one object to carry, the bag}. By using generating functions, we can use a single mathematical object to represent the complexity of the Bellman-Harris process. 

In this paper we explore and exploit the relationship between the Bellman-Harris process and the renewal model approach to infectious disease with the aim of clarifying and understanding the dynamical assumptions underpinning the renewal process. Particularly, we re-derive the generating function of the Bellman-Harris process describing $Z(t)$ with offspring distribution $q_n$. This leads to an integral equation, the expectation of which recovers the renewal formulation of epidemiology \cite{Fraser2007}. While this result in itself is not completely new (e.g. it was solved for binary offspring distributions in \cite{Bellman1948,Bellman1952}), we introduce two key generalisations. The first extends the Bellman-Harris process to allow for time-varying reproduction numbers, while the second provides a rigorous means of accounting for exogenous events, such as importations or zoonoses. 

Further, we show how our approach provides a flexible framework for extending both the observation models (i.e. the count noise around $E[Z(t)]$ and the effective reproduction number parametrisations, commonly employed in epidemic renewal model studies \cite{Cori2013, Nouvellet2018,Parag703751}. Specifically, we provide functionality for modelling negative binomial and Laplace distributed noise and for fitting autoregressive, polynomial and spline-based descriptions of the reproduction number. All source code is provided at \url{https://github.com/mrc-ide/bhrp}

\section{Deriving the general Bellman-Harris integral equation}
Part of the difficulty in understanding the Bellman-Harris equation is the list of objects needed to derive it and their assumptions. We start by listing them here for reference:

\begin{itemize}
    \item $Z(t)$ is a stochastic counting process for the number of infected individuals, $n\in\mathbb{I}^+$, existing at time $t\in \mathbb{R}^+$. We are trying to estimate the expectation $\mathbb{E}[Z(t)]$.
    \item  $\tau\in \mathbb{R}^+$ is a real random variable for the time taken to infect
    another individual. In infectious disease epidemiology, $\tau$ is generally called the serial interval distribution or offspring distribution\cite{Grassly2008}. $\tau$ has a probability measure, distributed by $\tau\sim g(\tau)$, with cumulative distribution function $G(\tau)$. A crucial assumption is that for each infected individual the distribution of $g$ is the same and independent of other infected individuals and examples of $g$ are shown in figure 1.
    \item $q_n$ is the probability that an individual infects $n$ other individuals. It is clear here that $q_n$ corresponds to the stochastic basic reproductive number and $q_n(t)$ to the stochastic time-varying reproductive number. The expected value of $q_n$ is the expected reproductive number that is ubiquitously used. 
    \item For an arbitrary variable $|s|\leq1,$ the generating function for the infection probabilities $\{q_n\}_{n}$ is: $h(s)=\sum_{n=0}^{+\infty} q_n s^n$. For infection probabilities to be time dependent $h(s,t)=\sum_{n=0}^{+\infty} q_n(t) s^n$. 
    \item We also introduce probabilities $p_r(t)$ of having $r$ infected individuals at time $t: p_r(t)=P(Z(t)=r).$ We can therefore write the generating function for an arbitrary variable $|s|\leq1$ and the number of infected individuals $Z(t)$ as: $F(s,t)=\sum_{r=1}^{\infty} p_r(t)s^r$.
\end{itemize}

Given the above expressions and the generating functions $h(s,t)$ and $F(s,t)$, we first need an expression for $p_r(t)=P[Z(t)=r]$, the probability that there are $r$ infected individuals at time $t$. Deriving an expression for $p_r(t)$ is challenging because there are many facets to evaluating this probability. To aid understanding we will explain how to arrive at an expression for $p_r(t)$ and give a simple example for the binary case, where each individual infects two others. The first consideration is that any infected individual could infect $n$ others (in time), and therefore we need to first sum over all possible infection probabilities $\sum_{n=0}^\infty q_n(t)$. The next consideration is that there are many ways to arrive at $p_r(t)$ and so we need to account for all the combinations of $n$ integers $\{i_1, i_2, \ldots, i_n \}$ such that we can have $r$ infections i.e. $i_1+i_2+\ldots+i_n=r$. Therefore, we want to sum over all the ways to get $r$ and multiply these probabilities: $\sum_{i_1+i_2+\ldots+i_n=r} \prod_{k=1}^n p_{i_k}(t-\tau)$. $\tau$ appears in this equation because the $r$ infections have occurred at some time $\tau$ before $t$. Finally, we need to integrate (average over) all the possible times $\tau$ at which the infections occurred, that is $\int_{\tau=0}^t \cdot dG(\tau)$. In the binary case, much simplifies and the first sum disappears because each individual always infects 2 others and the summation is not necessary i.e. $n \neq 2, q_n=0, \text{ and } q_2=1$. The second sum is also simply $\sum_{i_1+i_2=r}$ because two individuals are always infected. Therefore, if we wanted to know the probability of seeing say $4$ infections at time $t$ in the binary case, we would need to sum over $\{p_1(t-\tau)*p_3(t-\tau),p_3(t-\tau)*p_1(t-\tau),p_2(t-\tau)*p_2(t-\tau)\}$ and integrate $\tau$ with respect to the cumulative distribution function. In this binary case it is useful to notice that after the first infected individual infects two others, the process is self similar; a useful intuition therefore is to count taxa on a binary tree. Note that the Bellman Harris formulation is general and that if $G$ is a step function we have the Galton-Watson family tree model and if $G$ is exponentially distributed we have a Markov branching model. Putting all these constituents together, we can arrive at an expression for $p_r(t)$ that integrates (or averages) over incorporates the full uncertainty of when previous infections occurred and how many there were at that time. In what follows we will consider an extension of the original Bellman Harris formulation with $q_n(t)$, that is the probability of an individual infecting $n$ others varies with time such that 
\begin{equation}
    p_r(t)=\sum_{n=0}^{\infty}q_n(t) \Bigg(\sum_{i_1+i_2+\ldots+i_n=r} \int_{\tau=0}^{t} \prod_{k=1}^{n}p_{i_k}(t-\tau) dG(\tau) \Bigg).
\end{equation} 
Another intuition that may help the reader is that in the above equation, the terms in the braces are the coefficients of the generating function $h(s,t)$, which therefore have the standard interpretation of the probability of seeing $r$ infections while factoring all the ways $r$ can appear i.e. $p_r$.

Now that we have an expression for $p_r(t)$, we can arrive at the generating function $F(s,t)$ through multiplication with $s^r,$ over all possible values for $r$. Since we start with one infected individual, $r \geq 2$. Also we note that $s^r$ can be decomposed into its constituents $s^r=s^{i_1} s^{i_2} \ldots s^{i_n}:$
\begin{align}\label{sum_p}
 \sum_{r=2}^{\infty} p_r(t) s^r &= \sum_{r=2}^{\infty} \sum_{n=0}^{\infty}q_n(t) \Bigg(\sum_{i_1+i_2+\ldots+i_n=r} \int_{\tau=0}^{t} \prod_{k=1}^{n}p_{i_k}(t-\tau)
 dG(\tau) \Bigg)s^r\\
  &= \sum_{r=2}^{\infty} \sum_{n=0}^{\infty}q_n(t) \Bigg(\sum_{i_1+i_2+\ldots+i_n=r} \int_{\tau=0}^{t} \prod_{k=1}^{n}p_{i_k}(t-\tau)s^{i_k}
 dG(\tau) \Bigg)\\
 &= \int_{\tau=0}^{t} \sum_{n=0}^{\infty}q_n(t) \Bigg(\sum_{r=2}^{\infty} \Big( \sum_{i_1+i_2+\ldots+i_n=r} \prod_{k=1}^{n} p_{i_k}(t-\tau)s^{i_k} \Big) \Bigg) dG(\tau)\,.
\end{align}
The interchange of sums and integrals is a consequence of Fubini/Tonelli conditions (i.e all the above functions are positive and the integral with respect to $\tau$ converges absolutely).

To simplify the above equation, we can now refactor the individual constituents of $r$ - the indices $\{i_1+i_2+\ldots+i_n=r\}$. In this simplified notation, the fact that $r$ is made up of a multitude of infection branches is hidden to simplify subsequent derivation. However the reader should remember that $r$ is comprised of many possible branches and combinations.
\begin{equation}
\label{refactorisation}
    \sum_{r=2}^{\infty} \Big( \underbrace{ \sum_{i_1+i_2+\ldots+i_n=r} \prod_{k=1}^{n}  p_{i_k}(t-\tau)s^{i_k}} \Big) = \left(\sum_{r=2}^{\infty}  p_r(t-\tau) s^r \right)^n,
\end{equation} 
To see how equation \ref{refactorisation} is arises, note that:
\begin{align}
\Big(\sum_{r=0}^{\infty}p_r s^r\Big)^n
&=\sum_{r=0}^{\infty}\Big( \sum_{i_1+\ldots+i_n=r}p_{i_1}p_{i_2} \ldots p_{i_n} s^r \Big)\\
&=\sum_{r=0}^{\infty}\Big( \sum_{i_1+\ldots+i_n=r}p_{i_1}p_{i_2} \ldots p_{i_n} s^{i_1} s^{i_2}\ldots s^{i_n}\Big)\\
&=\sum_{r=0}^{\infty}\Big( \sum_{i_1+\ldots+i_n=r} \prod_{k=1}^{n} p_{i_k}s^{i^k}\Big).
\end{align}

so that the generating function $F(s,t)$ can be written more compactly as
\begin{equation}
\label{condensed_pr}
\sum_{r=2}^{\infty} p_r(t) s^r =  \int_{\tau=0}^{t}  \sum_{n=0}^{\infty}q_n(t) \left( \sum_{r=2}^{\infty}  p_r(t-\tau) s^r  \right)^n dG(\tau).
\end{equation} 

To simplify equation \ref{condensed_pr}, we need to remember several choices we have made and defined. First if we define the generating function of $Z(t)$ as $F(s,t)=\sum_{r=2}^\infty p_r(t)s^r,$ then clearly for some other time $t-\tau$, the generating function is:
\begin{equation}
F(s,t-\tau)=\sum_{r=2}^{\infty}  p_r(t-\tau)s^r.
\end{equation}
Finally, remembering that the generating function for infection probabilities is $h(s,t)=\sum_{n=0}^\infty q_n(t) s^n,$ we can rewrite: 
\begin{equation}
\label{pre_bellman}
\sum_{r=2}^{\infty} p_r(t) s^r =  \int_{\tau=0}^{t} \underbrace{\left( \sum_{n=0}^{\infty}q_n(t) \big(F(s,t-\tau) \big)^n \right)}_{h\left(F(s,t-\tau),t\right)} dG(\tau) = \int_{\tau=0}^t h\left(F(s,t-\tau),t\right)dG(\tau).
\end{equation}
Notice in this equation that the under brace can be simplified using the generating function $h$, $h(s,t)=\sum_{n=0}^{+\infty} q_n(t) s^n$. While this step can seem confusing, it is simply using the two generating functions and noticing recursive relationships. To finally arrive at the generating function $F(s,t)=\sum_{r=1}^{\infty} p_r(t)s^r$, we need to take equation \ref{pre_bellman} and sum from $r=1$. For $r=1$, $p_1(t)=1-G(t)$, or the probability of the index case \emph{not} infecting anyone. $F(1,t)$ is therefore $F(1,t)=(1-G(t))s^1$ from the definition of our generating function. Putting this all together we arrive at the celebrated Bellman-Harris integral equation:
\begin{equation}\label{generalBH}
    F(s,t)=\left(1-G(t)\right)s + \int_{\tau=0}^t h\left(F(s,t-\tau),t\right)dG(\tau).
\end{equation}
Normally the Bellman-Harris equation is derived with $q_n$ not $q_n(t)$ resulting in the more familiar
\begin{equation}\label{generalBH2}
    F(s,t)=\left(1-G(t)\right)s + \int_{\tau=0}^t h\left(F(s,t-\tau)\right)dG(\tau).
\end{equation}
We have derived both expressions to show how the expected basic reproductive number and time-varying reproductive numbers (a fundamental metric in infectious disease modelling) can arise. This will be shown when taking expectations below.

\section{The renewal equation}

At first glance the generating function in equations \ref{generalBH} and \ref{generalBH2} seems impenetrable and unsolvable. However, by exploiting the favorable properties of generating functions, we can easily calculate the moments of $Z(t)$ by calculating the derivatives of the generating function at $s=1$.

Remember, the generating function of the number of infected individuals at time $t$, $Z(t)$, is $\sum_{r=1}^{\infty}p_r(t)s^r$. We denote the first moment (mean) of the generating function as the first derivative evaluated at $s=1$:
\begin{equation}
F(s,t)= \mathbb{E}[s^{Z(t)}]=\sum_{r=1}^{\infty}p_r(t)s^r  
\end{equation}
\begin{equation}
\mathbb{E}[Z(t)] = f(t)= \frac{\partial F(s,t)}{\partial s}|_{s=1}= \sum_{r=1}^{\infty} r p_r(t).
\end{equation}

Here $f(t)$ is the average or expected number of infected individuals at time $t$. To get $f(t)$ we take the first derivative of the Bellman-Harris integral equation \eqref{generalBH}, and get: 
\begin{equation}\label{BH_1st_der}
    \frac{\partial F(s,t)}{\partial s}=1-G(t)+\int_{\tau=0}^{t}\frac{\partial F(s,t-\tau)}{\partial s} \frac{\partial h}{\partial s}\left(F(s,t-\tau),t\right)g(\tau) d\tau.
\end{equation}
Now, by using the properties of generating functions and evaluating at $s=1$ 
\begin{equation} \frac{\partial F(s,t)}{\partial s} |_{s=1}=f(t) = 1-G(t)+\int_{\tau=0}^{t} \underbrace{\frac{\partial F}{\partial s}(1,t-\tau)}_{f(t-\tau)} \frac{\partial h}{\partial s}\left(F(1,t-\tau),t\right)g(\tau) d\tau.
\end{equation}
From the definition of the generating function $F(1,t-\tau)=\sum_{r=1}^{\infty}p_r(t-\tau)1^r = \sum_{r=1}^{\infty}p_r(t-\tau)=1,$ and the derivative of the generating function $h(s,t)$, is $\frac{\partial h}{\partial s}(s,t)=\sum_{n=1}^{\infty}n q_n(t) s^{n-1}$ so to compute the average number of individuals a infected individual infects we again take the derivative of the generating function and evaluate at $s=1$,
\begin{equation}
   \frac{\partial h}{\partial s}\left(F(1,t-\tau),t\right)=\frac{\partial h}{\partial s}(1,t)= \sum_{n=1}^{\infty}n q_n(t)=R_t.
\end{equation}

This quantity is a the time-varying reproductive number. If we did this derivation using a fixed, time invariant $q_n$, then the expected value would have been
\begin{equation}
\frac{\partial h}{\partial s}\left(F(1,t-\tau)\right)=\frac{\partial h}{\partial s}(1)= \sum_{n=1}^{\infty}n q_n=R_0.
\end{equation}
These two equations show clearly how the basic and time-varying reproductive number arises from a carefully designed stochastic counting process. There is a clear intuition how reproductive numbers arise: first we design a time dependent branching process where there is a stochastic number, $n$, individuals that can be infected in time. Taking the expected value of $q_n$ from the first derivative of its generating function evaluated at $s=1$ gives us the expected reproductive numbers - or the average number of infections a given individual infects. In this paper we show that this can be made time varying by changing $q_n$ to $q_n(t)$.

Incorporating $R_t$ we get a renewal equation for the first moment of $Z(t)$:
\begin{equation}\label{renewal_m1}
    \mathbb{E}[Z(t)]= f(t) = 1-G(t)+R_t \int_{\tau=0}^{t} f(t-\tau)g(\tau) d\tau,
\end{equation}
where the term $1-G(t)$ is the survival probability of the first infection or index infection. This term arises from $p_1(t)$, or the probability of seeing the first infection at time $t$. 
\section{Imported or exogenous sources of infection}

In the renewal equation, there are two major components. The first term arises from $p_1(t)$, or the probability of seeing the first infection at time $t$. By assuming this first infection occurs straight away, this term disappears and we derive the renewal equation widely used in epidemic modelling \cite{Cori2013,Nouvellet2018}. Even when not making this assumption, the survival function tends to zero with time and therefore does not not contribute to the epidemic, but simply starts it. And once started, the second convolutional term in the renewal equation accounts for secondary cases and the chain of new infections - a convolution of its own history with some serial interval distribution. However, while the second term, that arises as a logical consequence from an age-dependent branching process, and does intuitively represent secondary infections in an outbreak, there is no term in the renewal equation for repeated seed infections or an exogenous source. There is only ever one seed infection or index case.

Exogenous sources are an important component in epidemic modelling, and account for infections entering from outside a system. Depending on the scale of modelling, it can represent zoonosis events, or importations from one geographic region to another by flights, etc. The splitting of the epidemic into exogenous and endogenous components, to our knowledge, is not formally defined in any previous renewal frameworks. From a intuitive view, not having these components separated means we only ever look at part of the epidemic. Consider the case of an influenza epidemic. A person flies in from a country experiencing an outbreak, this person then travels to a city where secondary infection occurs. The reproductive number and convolution part of the renewal equation describes this process, but \emph{does not} account for repeated introductions from other individuals flying in. Similarly, just looking at flights would only show part of the picture. This phenomenon is very general and observed in multiple disciplines \cite{Sornette2006}. 

\begin{figure}[htbp] \centering  \includegraphics[width=1\textwidth]{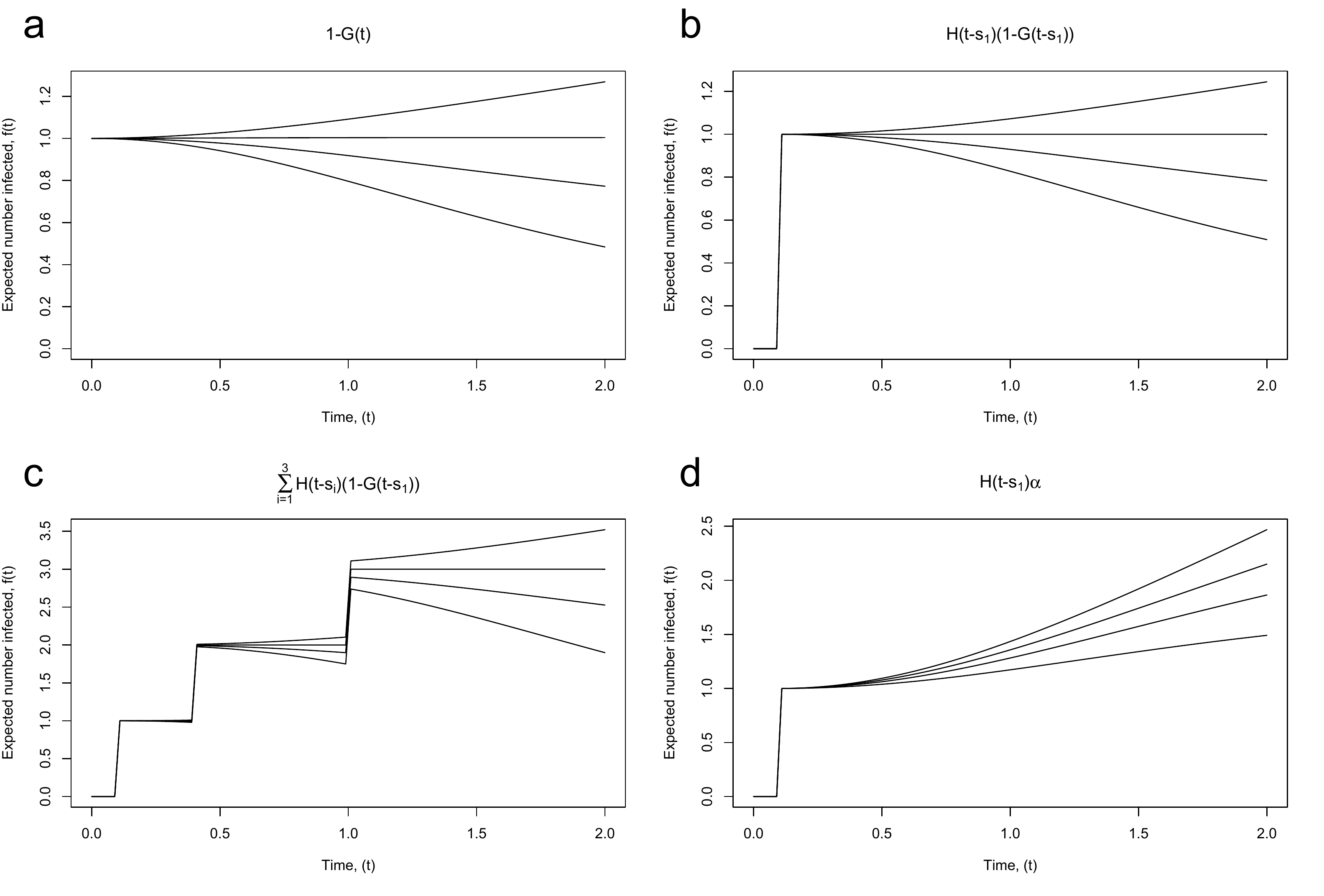}  \caption{\small $4$ different exogenous terms (a) $1-G(t)$ (b) $ \mathcal{H}(t-s_i)(1-G(t-s_i))$ (c) $\sum_{i=1}^{n}\mathcal{H}(t-s_i) (1-G(t-s_i))$ and (d) $ {H}(t-s_i)\alpha$, with $R_0=\{0.5,0.8,1,1.2\}$, with serial interval distribution from a Rayleigh(1)}. \label{Figure1} \end{figure}

In this section we will derive a renewal equation that accounts of exogenous sources at arbitrary times. It is useful to visualise the terms in the renewal equation (Figure 1) to see how it can easily be extended to include an exogenous component. Let us first consider a hypothetical scenario, or a new disease occurring in a homogeneously mixed immunologically naive population.  We say this new disease has a serial interval distribution that is Rayleigh ($g(\tau)=\sigma\tau e^{-\frac{1}{2}\sigma\tau^2}$) with $\sigma=1$. The renewal equation is
\begin{eqnarray}
    \label{condensed}
    f(t) &=& 1-G(t)+R_0  \int_{\tau=0}^t f(t-\tau)g(\tau) d\tau\\
     &=& \mu(t) + R_0 f * g(t)
\end{eqnarray}
where $*$ is a condensed convolution notation. We can solve the renewal equation using quadrature and see how the expected number of infected individuals evolves through time. Figure 1a shows the solution to equation \ref{condensed} for an the index case starting at $t=0$, and for $R_0=\{0.5,0.8,1,1.2\}$. However, equally, we can solve the renewal equation for a single index case starting at some future time $t=s_1$,
\begin{eqnarray}
    f(t) &=& (1-G(t-s_1))\mathcal{H}(t-s_1) + R_0 f * g(t)\\
     &=& \mu(t-s_1) + R_0 f * g(t),
\end{eqnarray}
where here we have introduced the Heaviside step function,
\begin{equation}
\mathcal{H}(x) \left\{\begin{matrix}
0,\quad x<s_i \\
1,\quad x>s_i \\ 
\end{matrix}\right.
\end{equation}
for importation time $s_i$.
Figure 1b shows this delayed importation, where the whole process shifts to start after $s_1$, after which it is identical to the unshifted version.  

And sum multiple renewal processes with exogenous inputs at different times i.e.:
\begin{eqnarray}
f(t) &=& \sum_{i=1}^{n}f_i(t)\\
&=& \sum_{i=1}^{n}\mathcal{H}(t-s_i)\left(1-G(t-s_i)\right) + R_0 f * g(t) \\
&=& \sum_{i=1}^n \mu(t-s_i) + R_0 f * g(t)
\end{eqnarray}
Here exogenous sources occur at times $s_i$, and the endogenous process begins at $s_1$. Our superposition of multiple renewal processes differs from the S-renewal process\cite{Teresalam1991} in that we do not consider differing convolutional terms. We justify this by saying there is no difference in the exogenous infections, they are the same disease as the endogenous ones, except they have entered the system from the outside and not as a secondary infection. To our knowledge this is a new derivation of the superposition renewal equation. To understand the dynamics of this superposition. In Figure 1c there are three exogenous events at times $s_1=0.1$, $s_2=0.4$, $s_3=1$. These exogenous infections seed the epidemic and allow it to increase. It is easy to see that an arbitrary function can be used in place of the Heaviside step function. As an example, figure 1d shows the example of a constant rate of importation at all times $t$, i.e.\ $\alpha(1-G(t-s_1))$. In this figure the number of infections does not reduce beyond the exogenous rate.  This  derivation justifies the use of an exogenous term in the renewal equation and specifies how to create one. 

There are two considerations to note, first, this exogenous term can be used to estimate when the first case occurred in an epidemic. Second, when performing inference, an analytic expression for the Heaviside  function such as $\frac{1}{1+e^{2kx}}$ for an arbitrary large $k$ can be used. Our final renewal equation is:
\begin{equation}\label{renewal_m1}
    \mathbb{E}[Z(t)]= f(t) = \underbrace{ \mu(t)}_{exogenous}+\underbrace{R_0 \int_{\tau=0}^{t} f(t-\tau)g(\tau) d\tau}_{endogenous}
\end{equation}
It is of interest here to note that equation \ref{renewal_m1} has deep connections with other counting processes such as Hawkes self-exciting stochastic processes \cite{Hawkes1971}. Indeed, in a nice convergence of theory, it has been shown that the expected value of the Hawkes intensity function results in exactly the same renewal equation we have derived \cite{Rizoiu2017}.

\section{Inference using the renewal equation}
The renewal equation \ref{renewal_m1} can be solved in closed form for a few special cases such as an exponential $G(t)$ (Markov branching process), and often the equation is studied in $\lim t\rightarrow \infty$. In such cases the Laplace transform is taken to simplify the convolution to a product. Numerically, quadrature can be used, which results in solving a system of linear equations or using approaches like the Trapezium rule. 

Often however, data is not continuous and binned into hourly, daily etc counts. For discrete data, the equation can be represented in binned form as:
\begin{equation}\label{renewal_binned}
     f(t)=\mu(t)+R_t \sum_{\tau<t} f(t-\tau)g(\tau)
\end{equation}
In this binned discrete form, the model for $f$ can be thought of as an Autoregression with coefficients determined by the serial interval distribution $g(\tau)$. Once again this is a convergence of theory, \cite{Rizoiu2017} show that the expectation of the intensity function of a Hawkes process results exactly in equation \ref{renewal_m1}, and \cite{Kirchner2015,Kirchner2017} have shown that a binned Hawkes process
is an AR($\infty$) process, that can be approximated by an AR($p$), with $p$ lags. Therefore the renewal equation, in discrete form, is therefore deeply connected to standard time series approaches. These time series approaches are very effective \cite{Makridakis2018} in forecasting but do not yield useful information about the underlying epidemiological mechanism and are largely "black box". Using the discrete renewal equation can have all the benefits of time series forecasting but is built from a mechanism rooted in infectious disease epidemiology and therefore has explainable and interpretable dynamics.

Equation \ref{renewal_binned} is highly flexible, and complicated non-parametric functions can be used for $\mu(t),$ and $R_t$. From a computational complexity view, equation \ref{renewal_binned} is quadratic $\mathcal{O}(n^2)$, which is limiting, but given the discretisation of time can handle most epidemic data. The equation is also easy to evaluate and is therefore amenable to implementation in state-of-the-art Bayesian MCMC software such as Stan. In a Bayesian Hierarchical framework a general model would be:
\begin{eqnarray}
\theta,\phi, \mu_t, R_t &\sim& p(\cdot) \\
 f(t) &=& \mu_t+R_t \sum_{\tau<t} f(t-\tau)g(\tau)\\
 y&\sim& p(f(t),\phi)
\end{eqnarray}
In this hierarchical Bayesian framework, the first line are the prior distributions for hyperparameters $\theta,\phi,\mu_t$ and $R_t$. $\theta$ is a hyperparameter for the serial interval distribution, and can be given vague or strong priors, or fixed (given known biology). $\phi$ is a parameter for variance or overdispersion. $\mu(t)$ is the exogenous component, that models the number of new infections entering the modelled system at time $t$. It can incorporate information on movement etc. $R_t$ is modelled by some function such as a stochastic process, polynomials, splines etc. The second line is the discrete binned renewal equation. We note here, $g(\tau)$ can be multivariate and incorporate information on genetics or spatial distance. Typically, $g(\tau)$ is a Rayleigh, Log Normal, or Gamma distribution\cite{Cori2013}. Equivalently, $\mu(t)$ can spatially vary and can be modelled effectively by log Gaussian Cox Processes. The third and final line is the likelihood function, which for aggregated count data is generally negative binomial, i.e.
\begin{equation}
    y\sim\binom{y+\phi-1}{y}\left(\frac{f(t)}{f(t)+\phi}\right)^y\left(\frac{\phi}{f(t)+\phi}\right)^\phi.
\end{equation}
Our negative binomial is the reparameterisation where the location parameter, or mean, is our renewal equation $E[y]=f(t)$ and the variance is $E[(y-E[y])^2]=f(t)+\frac{f(t)^2}{\phi}$.
The unnormalised posterior distribution is then 
\begin{equation}
\label{posterior}
  p(\theta,\phi,\mu_t,R_t|y) \propto p(y|f(t),
\phi)p(f(t)|\theta,\mu_t,R_t)p(\theta)p(\mu_t)p(R_t)p(\phi). \end{equation}
Posterior expectations can be calculated through approximate inference or via full MCMC sampling.
\newpage

\section{Fitting a renewal model to South Korea data from the COVID-19 pandemic}

\begin{figure}[htbp] \centering  \includegraphics[width=1\textwidth]{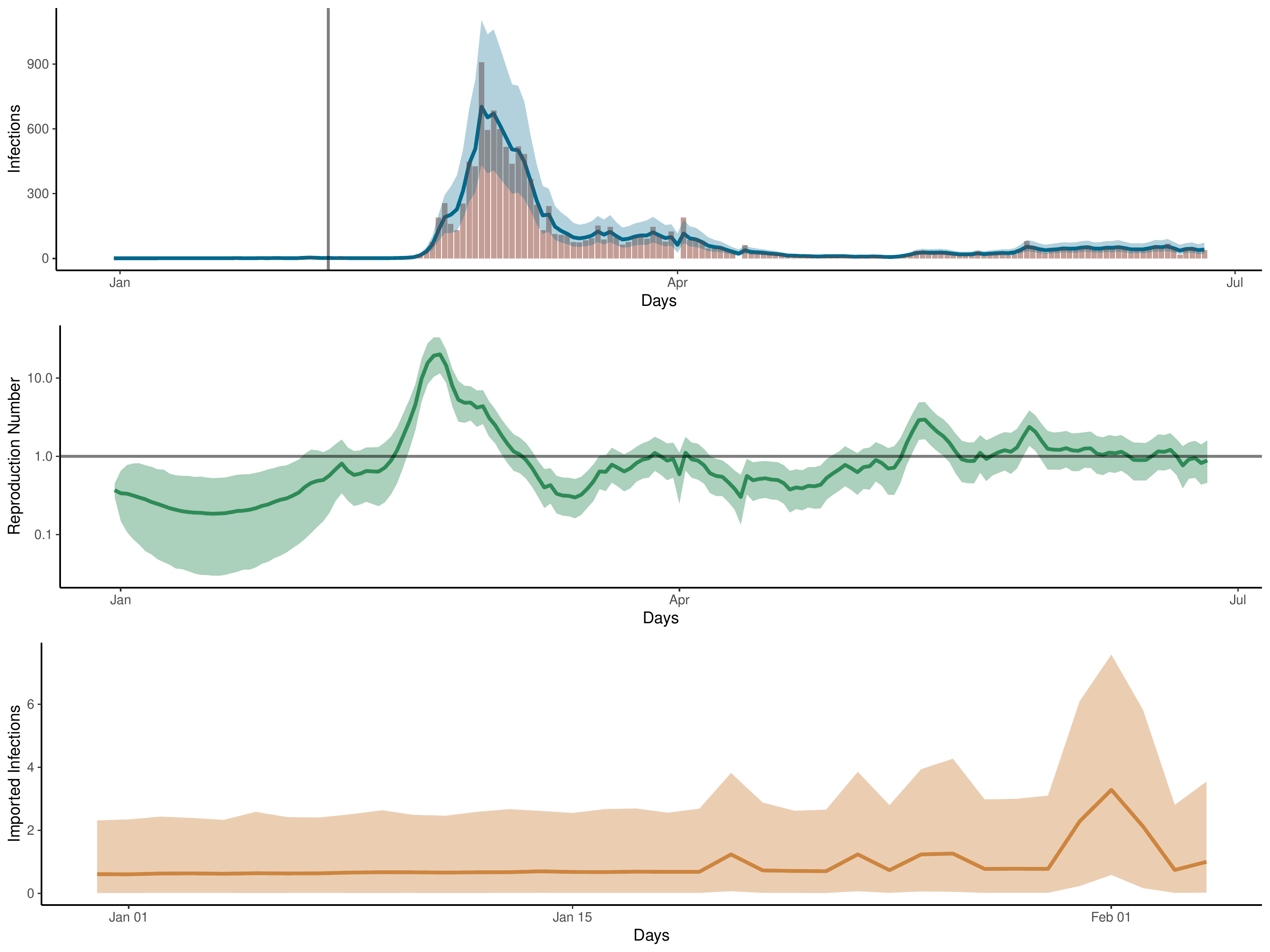}  \caption{Top: Observed cases (red), predicted cases (blue line), and 95\% credible interval (blue ribbon). Middle: Estimated reproduction number (green line) with 95\% credible interval (green ribbon). Bottom: Estimated importations (orange line) with 95\% credible interval (orange ribbon). } \label{Figure2} \end{figure}

We fit a discrete renewal process with an exogenous component to COVID-19 case data for South Korea. Given the high testing capacity in South Korea, case data is generally considered reliable. However extensions can be included to model death data \cite{Flaxman2020}. We consider $g(\tau)$ to be the serial interval distribution and fixed as Gamma\cite{Bi2020} i.e:
\begin{equation*}
    g \sim \text{Gamma}(6.5,0.62).
\end{equation*}
The serial interval distribution is discretised as $g_s=\int_{s-0.5}^{s+0.5} g(\tau) d\tau$ for $s=2,3,...,$ and $g_1 = \int_{0}^{1.5} g(\tau)d\tau$. We paramaterise the time-varying reproduction number as $R_{t} = exp(\epsilon_t) $, with the exponential to ensure positivity. $\epsilon_t$ is a AR(2) process that starts with $\epsilon_{1}\sim N(-1,0.1)$, and  $\epsilon_{2}\sim N(-1,\sigma_t^\ast)$:
\begin{equation}
\label{def:weeklyAR}
\epsilon_{t}\sim N(\rho_1\epsilon_{t-1} +\rho_2\epsilon_{t-2}, \sigma_t^*)\, \text{for }t=\{3,4,5,\dots\},
\end{equation} 
with independent  priors on $\rho_1$ and $\rho_2$ that are normal distributions conditioned to be in $[0,1]$; the prior for $\rho_1$ is a $N(0.8,.05)$ distribution constrained to the unit interval and  the prior for  $\rho_2$\ is a $N(0.1,0.05)$ distribution constrained to the unit interval. The prior for $\sigma_t$, the standard deviation of the stationary distribution of $\epsilon_t$ is chosen as $\sigma_t\sim N^+(0,1)$. The standard deviation of the weekly updates to achieve this standard deviation of the stationary distribution is ~$\sigma_t^\ast=\sigma_t\sqrt{1-\rho_1^2-\rho_2^2-2\rho_1^2\rho_2/(1-\rho_2)}$. Other stochastic processes such as a Gaussian process or random walks could also be readily used. The exogenous component, accounting for importations into South Korea was modelled for the first $t=\{1,\dots,40\}$ days, from 2019/12/31 upto 2020/02/04 where travel restrictions were imposed \cite{Hale2020}. We choose independent daily effects for $\mu_t\sim \text{Exponential}(0.5)$. The posterior distribution is then found through equation \ref{posterior}. Estimates from this model are shown in Figure \ref{Figure2}. For South Korea, there was a rapid rise in cases with a large $R_t$ in early January, followed by a reduction and a period of $R_t$ below 1.0. During this period, the cases numbers reduced so there were some days where no new cases were observed. In June, South Korea started experiencing smaller second waves with $R_t$ greater than 1.0 for short periods of time. Throughout this time our importation model fits a small but non-zero number of expected importations.

\section{Conclusion}

In this paper we have derived from first principles how the renewal equation arises from a stochastic age-dependent branching process. In this derivation we show how the time-varying reproduction number emerges, and how it is possible to incorporate the exogenous infections such as importations or zoonoses. We then highlight how to perform full Bayesian inference over the renewal equation and provide an example modelling COVID-19 case data in South Korea. 

The renewal equation is deeply connected with other approaches in epidemiology such as SEIR models \cite{Champredon2018}, Hawkes processes \cite{Rizoiu2017}, and autoregressive processes \cite{Kirchner2015}. The benefit of the more complicated derivation we have showcased in this paper is that it allows the ability to disentangle what is possible to model in the renewal equation by referring to the underlying assumptions in the age-dependent branching process. For example, it does not seem principled, within the renewal equation,  to allow the serial interval distribution to be time varying. To do so would result in a very different generating function. Rather, the serial interval distribution is interpreted as a fundamental property of a given pathogen; changes in the rate of transmission happen through the time-varying reproduction number.

Our derivation provides a means to incorporate more complex epidemiological processes within the renewal process but ensuring that these additions can be considered from a principled mathematical foundation.

\newpage

\section{Funding}
SB would like to acknowledge the NIHR BRC Imperial College NHS Trust Infection and COVID themes the Academy of Medical sciences Springboard award and the Bill and Melinda Gates Foundation. 

\printbibliography

\end{document}